\documentclass[12pt]{article}
\usepackage{epsfig}

\def\beq{\begin{equation}}
\def\eeq#1{\label{#1}\end{equation}}
\def\eeqn{\end{equation}}

\def\beqa{\begin{eqnarray}}
\def\eeqa#1{\label{#1}\end{eqnarray}}
\def\eeqan{\end{eqnarray}}

\let\bar=\overbar

\def\Dslash{\not{\hbox{\kern-4pt $D$}}}
\def\dslash{\not{\hbox{\kern-2pt $\del$}}}

\def\msb{{\bar{\ssstyle M \kern -1pt S}}}

\usepackage{fancyhdr,graphicx}
\def\Title#1{\begin{center} {\Large {\bf #1} } \end{center}}

\begin{document}

\Title{Strange stars with different quark mass scalings}

\bigskip\bigskip


\begin{raggedright}

{\it Ang Li\index{Li, A.}\\
Department of Physics and \\
Institute of Theoretical Physics
and Astrophysics\\
Xiamen University\\
Xiamen 361005\\
P. R. China\\
{\tt Email: liang@xmu.edu.cn}}
\bigskip\bigskip
\end{raggedright}

\section{Introduction}

In studying the equation of state (EOS) of ordinary quark matter,
the cruial point is to treat quark confinement in a proper way.
Except the conventional bag mechanism (where quarks are
asymptotically free within a large bag), an alternative way to
obtain confinement is based on the density dependence of quark
masses, then the proper variation of quark masses with density would
mimic the strong interaction between quarks, which is the basic idea
of the quark mass-density-dependent model.

Originally, the interaction part of the quark masses was assumed to
be inversely proportional to the density~(Fowler et al. 1981;
Chakrabarty 1991; Chakrabarty 1993; Chakrabarty 1996), and this
linear scaling has been extensively applied to study the properties
of strange quark matter (SQM). However, this class of scaling is
often criticized for its absence of a convincing derivation~(Peng
2000). Then a cubic scaling was derived based on the in-medium
chiral condensates and linear confinement~(Peng 2000). and has been
widely used afterwards~(Lugones $\&$ Horvath 2003; Zheng et al.
2004; Peng et al. 2006; Wen et al. 2007; Peng et al. 2008). But this
deriving procedure is still not well justified since it took only
the first order approximation of the chiral condensates in medium.
Incorporating of higher orders of the approximation would
nontrivially complicate the quark mass formulas~(Peng 2009). In
fact, there are also other mass scalings in the literatures~(Dey et
al. 1998; Wang 2000; Zhang et al. 2001; Zhang $\&$ Su 2002; Zhang
$\&$ Su 2003).

Despite the big uncertainty of the quark mass formulas, this model,
after all, is no doubt only a crude approximation to QCD. For
example, the model may not account for quark system where realistic
quark vector interaction is non-ignorable. However, we can not get a
general idea of how the strong interaction acts from the fundamental
theory of strong interactions in hand, i.e. QCD. Until this
stimulating controversy is solved, we feel safe to take the
pragmatic point of view of using the model. This work does not claim
to answer how Nature works. However, it may shed some light on what
may happen in interesting physical situations. In this respect, the
quark mass-density-dependent model has been, and still is, an
interesting laboratory.

The aim of the present paper then, is to study in what extent this
scaling model is allowed to study the properties of SQM. To this
end, we treat the quark mass scaling as a free parameter, to
investigate the stability of SQM and the variation of the predicted
properties of the corresponding strange stars (SSs) within a wide
scaling range. Furthermore, we try to demonstrate the general
features of SSs related to astrophysics observations, whatever the
value of the free parameters.

The paper is organized as follows. In Section 2 we describe the
formalism applied in calculating the EOS of the SQM in the quark
mass-density-dependent model. In Section 3 we present the structure
of the stars made of this matter, including mass-radius relation,
spin frequency, electric properties of the quark surface. Finally in
Section 4 we address our main conclusions.

\section{The Model}

As usually done, we consider SQM as a mixture of interacting $u$,
$d$, $s$ quarks, and electrons, where the mass of the quarks $m_{q}$
($q = u, d, s$)  is parametrized with the baryon number density
$n_{\mathrm{b}}$ as follows:
\begin{equation}
m_q \equiv m_{q0}+ m_{\mathrm{I}}=m_{q0}+\frac{C}{n_{\mathrm{b}}^x},
\label{mqT0}
\end{equation}
where $C$ is a parameter to be determined by stability arguments.
The density-dependent mass $m_{q}$ includes two parts: one is the
original mass or current mass $m_{q0}$, the other is the interacting
part $m_{\mathrm{I}}$. The exponent of density $x$, i.e. the quark
mass scaling, is treated as a free parameter in this paper.

Denoting the Fermi momentum in the phase space by $\nu_i$ ($i=u, d,
s,e^-$), the particle number densities can then be expressed as
\begin{equation} \label{nimod}
n_i =g_i\int \frac{\mathrm{d}^3{\bf p}}{(2\pi\hbar)^3}
=\frac{g_i}{2\pi^2} \int_0^{\nu_i}\, p^2\,\mbox{d}p
=\frac{g_i\nu_i^3}{6\pi^2}{\bf ,}
\end{equation}
and the corresponding energy density as
\begin{equation} \label{Emod}
\varepsilon
=\sum_i\frac{g_i}{2\pi^2}\int_0^{\nu_i}\sqrt{p^2+m_i^2}\,p^2\,\mbox{d}p{\bf
.}
\end{equation}

The relevant chemical potentials $\mu_u$, $\mu_d$, $\mu_s$, and
$\mu_e$ satisfy the weak-equilibrium condition (we assume that
neutrinos leave the system freely):
\begin{eqnarray}
 \mu_u+\mu_e=\mu_d, ~~~\mu_d=\mu_s{\bf .} \label{weak}
\end{eqnarray}

For the quark flavor $i$ we have
\begin{eqnarray}
\mu_i &=& \frac{\mathrm{d} \varepsilon}{\mathrm{d} n_i}
|_{\{n_{k\neq i}\}}= \frac{\partial \varepsilon_i}{\partial \nu_i}
\frac{\mathrm{d}\nu_i}{\mathrm{d} n_i}
 +\sum_j \frac{\partial \varepsilon}{\partial m_j}\frac{\partial m_j}{\partial n_i}
\nonumber \\
&=&
  \sqrt{\nu_i^2+m_i^2}
  +\sum_j n_j\frac{\partial m_j}{\partial n_i}
   f\!\left(\frac{\nu_j}{m_j}\right),
 \label{mui}
\end{eqnarray}
where
\begin{equation}
f(a) \equiv \frac{3}{2a^3} \left[
 a\sqrt{1+a^2}-\ln\left(a+\sqrt{1+a^2}\right)
\right].
\end{equation}
We see clearly from Equ.~(\ref{mui}) that since the quark masses are
density dependent, the derivatives generate an additional term with
respect to the free Fermi gas model.

For electrons, we have
\begin{equation} \label{muevsne}
\mu_e=\sqrt{\left(3\pi^2n_e\right)^{2/3}+m_e^2}{\bf .}
\end{equation}

The pressure is then given by
\begin{eqnarray}
P&=& -\varepsilon + \sum_i \mu_i n_i
\nonumber \\
&=&
  -\Omega_0
  +\sum_{ij} n_i n_j\frac{\partial m_j}{\partial n_i}
   f\left(\frac{\nu_j}{m_j}\right)
\nonumber \\
&=& -\Omega_0
  +n_{\mathrm{b}}\frac{\mathrm{d}m_{\mathrm{I}}}{\mathrm{d}n_{\mathrm{b}}}
   \sum_{j=u,d,s} n_j\ f\!\left(\frac{\nu_j}{m_j}\right){\bf ,}
 \label{pressure}
\end{eqnarray}
with $\Omega_0$\ being the free-particle contribution:
\begin{eqnarray}
\Omega_0 &=& -\sum_i\frac{g_i}{48\pi^2} \left[
 \nu_i\sqrt{\nu_i^2+m_i^2}\left(2\nu_i^2-3m_i^2\right)
\right.
\nonumber\\
&& \phantom{-\sum_i\frac{g_i}{48\pi^2}[}
 \left.
 +3m_i^4\,\mbox{arcsinh}\left(\frac{\nu_i}{m_i}\right)
\right].
\end{eqnarray}

The baryon number density and the charge density can be given as:
\begin{equation} \label{qmeq3}
n_{\mathrm{b}}=\frac{1}{3}(n_u+n_d+n_s){\bf ,}
\end{equation}

\begin{equation} \label{qmeq4}
Q_{\mathrm{q}}=\frac{2}{3}n_u-\frac{1}{3}n_d-\frac{1}{3}n_s-n_e.
\end{equation}
The charge-neutrality condition requires $Q_{\mathrm{q}}=0$.

Solving Equs. (\ref{weak}), (\ref{qmeq3}), (\ref{qmeq4}), we can
determine $n_u$, $n_d$,  $n_s$, and $n_e$ for a given total baryon
number density $n_{\mathrm{b}}$. The other quantities are obtained
straightforwardly.

In the present model, the parameters are: the electron mass
$m_e=0.511$ MeV, the quark current masses $m_{u0}$, $m_{d0}$,
$m_{s0}$, the confinement parameter $C$ and the quark mass scaling
$x$. Although the light-quark masses are not without controversy and
remain under active investigations, they are anyway very small, and
so we simply take $m_{u0}=5$ MeV, $m_{d0}=10$ MeV. The current mass
of strange quarks is $95\pm 25$ MeV according to the latest version
of the Particle Data Group~\cite{Yao06}

\begin{figure}[htb]
\begin{center}
\epsfig{file=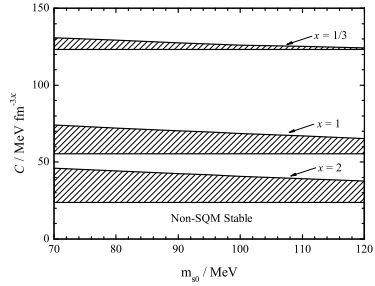,height=1.5in} \caption{The stability window
of the SQM at zero pressure with the quark mass scaling parameter $x
= 1/3, 1, 2$. The stability region (shadow region), is where the
energy per particle is lower than 930 MeV and two-flavor quark
matter is unstable.} \label{fig1}
\end{center}
\end{figure}

We now need to establish the conditions under which the SQM is the
true strong interaction ground state. That is, we must require, at
$P=0, E/A\leq M(^{56}{\rm Fe})c^2/56=930$ MeV for the SQM and
$E/A>930$ MeV for two-flavor quark matter (where $M(^{56}{\rm Fe})$
is the mass of $^{56}{\rm Fe}$) in order not to contradict standard
nuclear physics. The EOS will describe stable SQM only for a set of
values of ($C,m_{s0}$) satisfying these two conditions, which is
given in Fig.~1 as the ``stability window''. Only if the
$(C,m_{s0})$ pair is in the shadow region, SQM can be absolutely
stable, therefore the range of $C$ values is very narrow for a
chosen $m_{s0}$ value.

We then illustrate in Fig.~2 the density dependence of
$m_{\mathrm{I}}$ with the quark mass scaling $x = 1/10, 1/3, 1, 3$.
The calculation is done with $m_{s0}=95$ MeV and $C$ values
corresponding to the upper boundaries defined in Fig. 1 (the same
hereafter), that is, the system always lies in the same binding
state (for each $x$), i.e, E/A = 930 MeV. We presented those $C$
values in the last row of the Table.~1. Clearly the quark mass
varies in a very large range from very high density region
(asymptotic freedom regime) to lower densities, where confinement
(hadrons formation) takes place. It is compared with Dey et al.'s
scaling (dash-dotted)~\cite{Dey98}.

\begin{figure}[htb]
\begin{center}
\epsfig{file=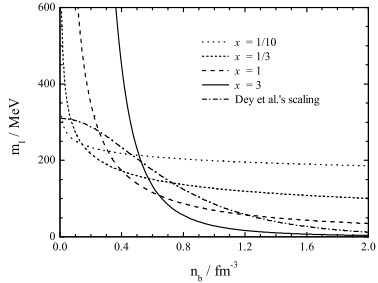,height=1.5in} \caption{The density
dependence of $m_{\mathrm{I}}$ with the quark mass scaling parameter
$x = 1/10,1/3, 1, 3$. The calculation is done with $m_{s0}=95$ MeV
and $C$ values presented in the last row of the Table.~1 (see text
for details). It is compared with Dey et al.'s scaling (dash-dotted
line)~(Dey et al. 1998).} \label{fig2}
\end{center}
\end{figure}
\section{Results and Discussion}

The resulting EOSs for the SQM are shown in the left panel of Fig.~3
with $x = 1/10, 1/5, 1/3, 1, 2, 3$. Because the sound velocity $v =
\mid dP/d\rho\mid^{~1/2~}$ should be smaller than $c$ (velocity of
light), unphysical region excluded by this condition has been
displayed with scattered dots. For the $x$ values chosen here, they
have quite different behavior at low density, basically falling into
two sequences. At small scalings ($x = 1/10, 1/5, 1/3$) the pressure
increases rather slowly with density; while the curve turns to
rapidly increase with density at relatively large $x$ values ($x =
1, 2, 3$). They cross at $\varepsilon \sim $ 800 MeV fm $^{-3}$,
then tend to be asymptotically linear relations at higher densities,
and a larger $x$ value leads to a stiffer EOS.

This behavior of EOSs would be mirrored at the prediction of
mass-radius relations of the corresponding SSs, as is shown in the
right panel of Fig.~3. For the first sequence, the maximum mass
occurs at a low central density (as shown in Table.~\ref{Table1}),
so a higher maximum mass is obtained due to a stiffer EOS, and with
the increase of $x$ value, the maximum mass is reduced from
1.78$M_{\odot}$ at $x = 1/10$ down to 1.61 $M_{\odot}$ at $x = 1/3$;
While we observe a slight increase of the maximum mass with $x$
value for the second sequence: from 1.56$M_{\odot}$ at $x = 1$ up to
1.62 $M_{\odot}$ at $x = 3$. Anyway the resulting maximum mass lies
between 1.5$M_{\odot}$ and 1.8$M_{\odot}$ for a rather wide range of
$x$ value chosen here (0.1 -- 3), which may be a pleasing feature of
this model: well-controlled. To see the region of stellar parameters
allowed by this model, we plot in Fig.~3 also the M(R) curves for
lower boundaries defined in Fig.~1 with $x = 1/5, 1/3, 1$ (grey
lines in the right panel).

Moreover, the radii invariably decrease with $x$ value. In addition,
we employ the empirical formula connecting the maximum rotation
frequency with the maximum mass and radius of the static
configuration~\cite{Gou99}, and present also the maximum rotational
angular frequency $\Omega_{\rm max}$ as $7730{\Large \left(
\frac{M_{\rm max}^{\rm stat}}{M_{\rm max}}
\right)^{\frac{1}{2}}\left( \frac{R^{\rm stat}_{M _{\rm max}}}{10
{\rm km}} \right)^{-\frac{3}{2}}}$rad s$^{-1}$. As a result, a
larger $x$ value results in a larger maximum spin frequency, SSs
with $x = 3$ can rotate at a frequency of 2194 rad s$^{-1}$. More
detailed results can be found in Table.~\ref{Table1}.

\begin{table}[b]
\begin{center}
 \begin{tabular}{cccccccccc}
  \hline
$x$ & $1/10$  &  $1/3$ & $1/2$ & $1$ & $2$ & $3$\\
  \hline
$M/M_{\odot}$  & 1.78  & 1.61 & 1.58 & 1.56 & 1.61 & 1.62\\
$R/{\rm km}$         & 13.2  & 9.38 & 8.75 & 8.10 & 7.97 & 7.89\\
$n_c/n_0$      & 4.35  & 7.88 & 8.88 & 10.1 & 10.2 & 10.3\\
$\Omega_{\rm max}/{\rm rad~s}^{-1}$      & 1066 & 1691 & 1860 & 2072 & 2159 & 2194\\
\hline
$C/{\rm MeV~fm}^{-3x}$& 199.1 & 126.8 & 104.1 & 69.5 & 41.7 & 28.8 \\
  \hline
\end{tabular}
\caption{Calculated results for the gravitational masses, radii,
central baryon densities (normalized to the saturation density of
nuclear matter, $n_0$ = 0.17 fm$^{-3}$), and the maximum rotational
frequencies for the maximum-mass stars of each strange star
sequence. The calculation is done with $m_{s0}=95$ MeV and
 $C$ values presented in the last row of this table.} \label{Table1}
\end{center}
\end{table}

\begin{figure}
\includegraphics[width=8.2cm]{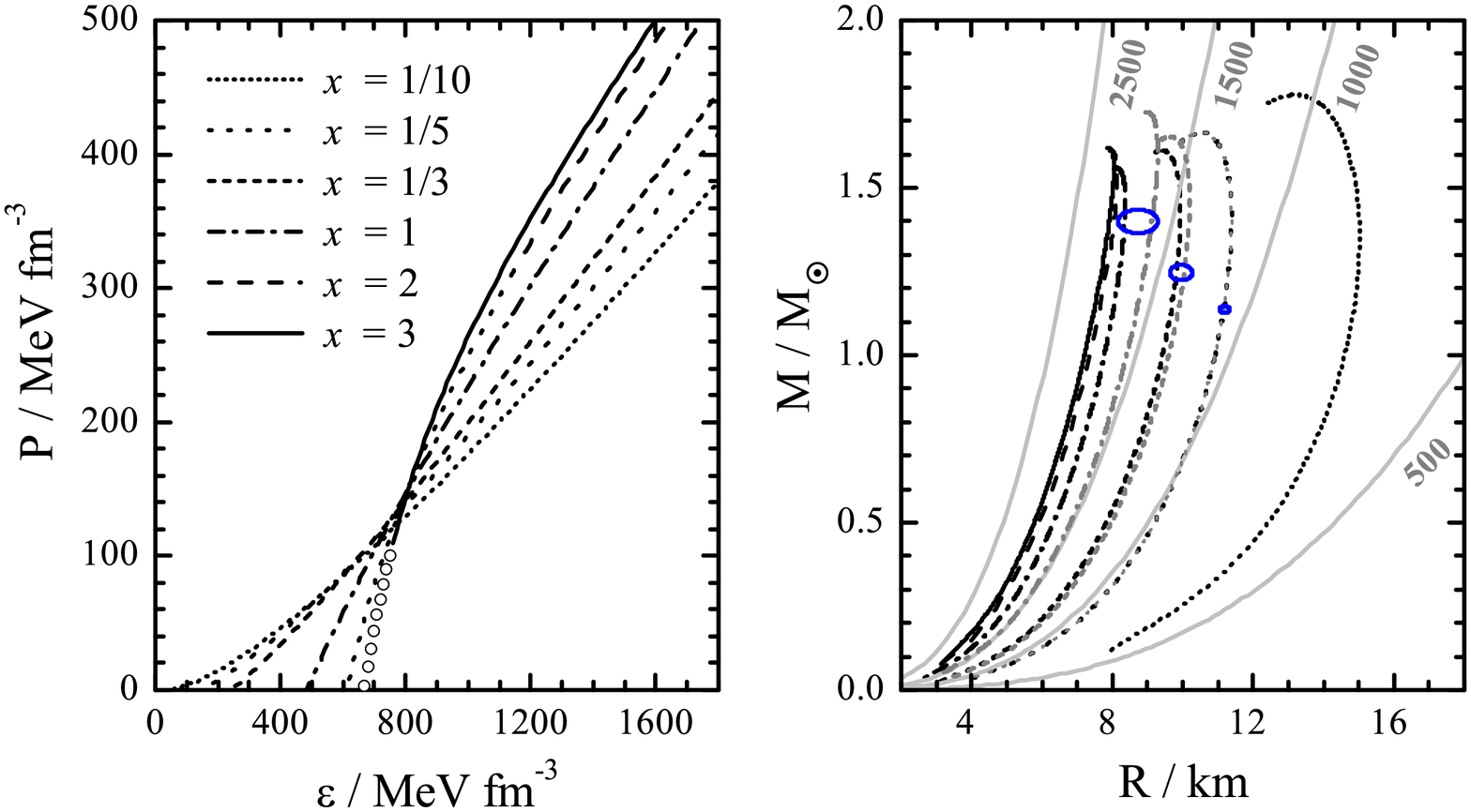}
\caption{The SQM EOS and mass-radius relation of SSs with $x = 1/10,
1/5, 1/3, 1, 2, 3$. M(R) curves for lower boundaries defined in
Fig.~1 with $x = 1/5, 1/3, 1$ are also presented (grey lines in the
right panel). Contours of the maximum rotation frequencies are given
by the light grey curves~(Gourgoulhon et al. 1999). } \label{fig3}
\end{figure}

In addition, the surface electric field should be very strong near
the bare quark surface of a strange star because of the mass
difference of the strange quark and the up (or down) quark, which
could play an important role in producing the thermal emission of
bare strange stars by the Usov mechanism~(Usov 1998; Usov 2001).
Moreover, the strong electric field plays an essential role in
forming a possible crust around a strange star, which has been
investigated extensively by many authors~(for a recent development,
see Zdunik et al. 2001). Also it should be noted that this electric
field has some important implications on pulsar radio emission
mechanisms~(Xu et al. 2001). Therefore it is very worthwhile to
explore how the mass scaling influences the surface electric field
of the stars, and possible related astronomical observations in turn
may drop a hint on what the proper mass scaling would be. Adopting a
simple Thomas-Fermi model, one gets the Poisson's equation~(Alcock
et al. 1986):
\begin{equation}
{d^2 V\over dz^2} = \left\{    \begin{array}{ll}
{4\alpha\over 3\pi}(V^3-V_{\rm q}^3) & z\leq 0,\\
{4\alpha\over 3\pi} V^3 & z > 0,
\end{array}     \right.
\end{equation}
where $z$ is the height above the quark surface, $\alpha$ is the
fine-structure constant, and $V_{\rm q}^3/(3\pi^2 \hbar^3 c^3)$ is
the quark charge density inside the quark surface.
Together with the physical boundary conditions $\{ z \rightarrow
-\infty: V \rightarrow V_{\rm q}, dV/dz \rightarrow 0;~~ z
\rightarrow +\infty: V \rightarrow 0,   dV/dz \rightarrow 0 \}$, and
the continuity of $V$ at $z=0$ requires $V(z=0) = 3V_{\rm q}/4$, the
solution for $z > 0$ finally leads to
\begin{equation}
V={3V_{\rm q}\over \sqrt{6\alpha\over\pi}V_{\rm q}z+4}~~ ({\rm
for}~z > 0).
\end{equation}
The electron charge density can be calculated as $ V^3/(3\pi^2
\hbar^3 c^3) $, therefore the number density of the electrons is
\begin{equation}
n_{\rm e}  =  {9V_{\rm q}^3\over \pi^2 (\sqrt{6\alpha\over\pi}V_{\rm
q}z+4)^3} \label{ne}
\end{equation}
and the electric field above the quark surface is finally
\begin{equation}
E = \sqrt{2\alpha\over 3\pi} \cdot {9 V_{\rm q}^2 \over
    (\sqrt{6\alpha\over \pi} V_{\rm q} \cdot z + 4)^2} \label{E}
\end{equation}
which is directed outward.

It is shown in Fig.~4 (take $x = 1/3$ for example) that although the
electric field near the surface is about $ 10^{18}$ V cm$^{-1}$, the
outward electric field decreases very rapidly above the quark
surface, and at $z\sim 10^{-8}$ cm, the field gets down to $\sim
5\times 10^{11}$ V cm$^{-1}$, which is of the order of the
rotation-induced electric field for a typical Goldreich-Julian
magnetosphere. To alter the mass scaling mainly has two effects:
First, it affects a lot the surface electric field, and a small
scaling parameter leads to an enhanced electric field. The change of
electric field would be almost a order of magnitude large (from
$10^{17}$ V cm$^{-1}$ to $10^{18}$ V cm$^{-1}$), which may have some
effect on astronomical observations. Second, a larger scaling would
slow the decrease of the electric field above the quark surface.

\begin{figure}[htb]
\begin{center}
\epsfig{file=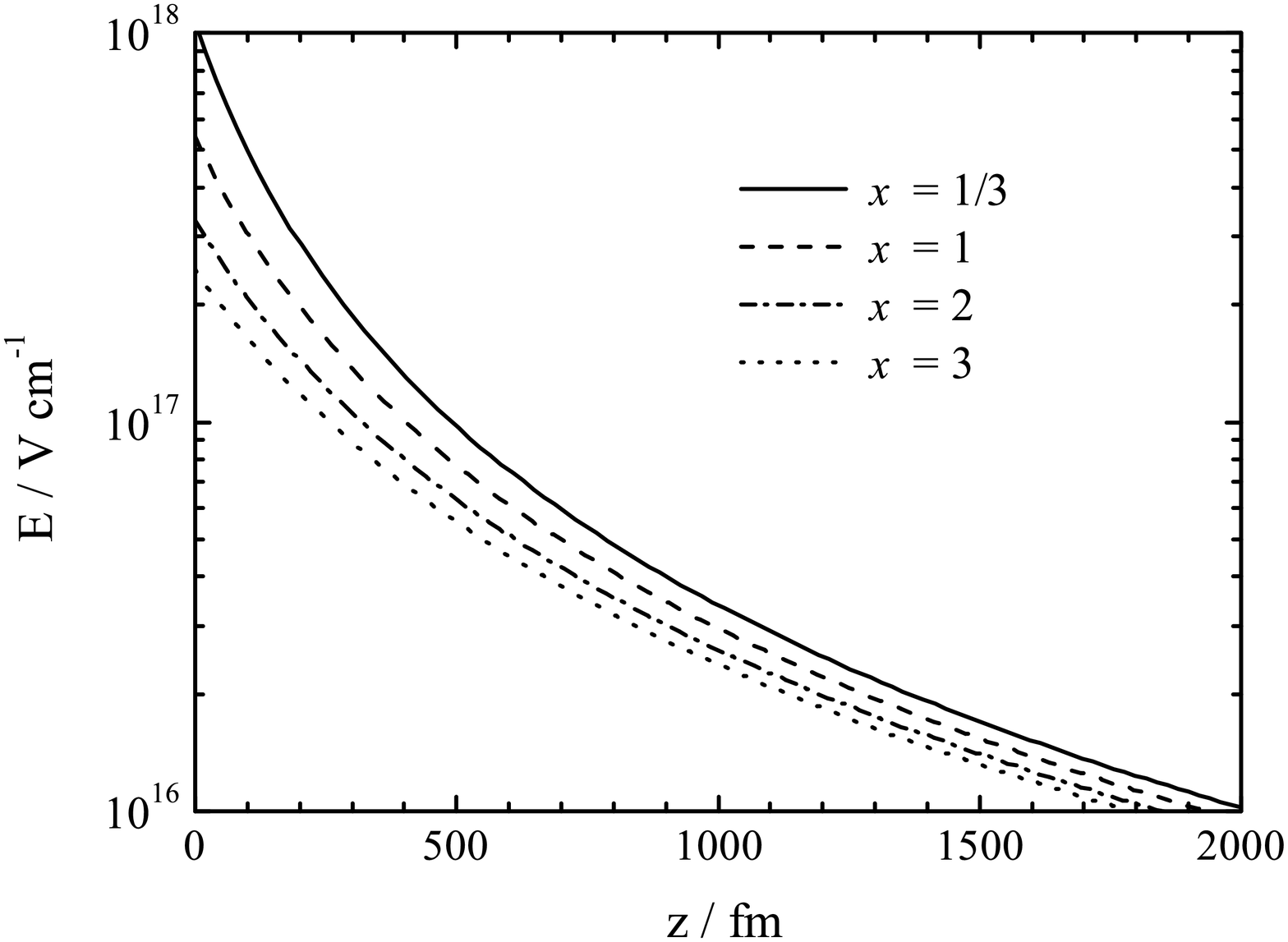,height=1.5in} \caption{The electric field
above the quark surface with $x =  1/3, 1, 2, 3$.} \label{fig4}
\end{center}
\end{figure}
\section{Conclusions}

In this paper, we investigate the stability of SQM within a wide
scaling range, i.e. from 0.1 to 3. We study also the properties of
the SSs made of the matter. The calculation shows that the resulting
maximum mass always lies between 1.5$M_{\odot}$ and 1.8$M_{\odot}$
for all the mass scalings chosen here. Strange star sequences with a
linear scaling would support less gravitational mass, a change
(increase or decrease) of the scaling parameter around the linear
scaling would result in a higher maximum mass. Radii invariably
decrease with the mass scaling; and then the larger the scaling, the
faster the star rotates. In addition, the variation of the scaling
may cause an order of magnitude change of the surface electric
field, which may have some effect on astronomical observations.

\section*{Acknowledgments}

We would like to thank an anonymous referee for valuable comments
and suggestions, and acknowledge Dr. Guang-Xiong Peng for beneficial
discussions. This work was supported by the National Basic Research
Program of China under grant 2009CB824800, the National Natural
Science Foundation of China under grants 10778611 and 10833002, and
the Youth Innovation Foundation of Fujian Province under grant
2009J05013.

\def\Discussion{
\setlength{\parskip}{0.3cm}\setlength{\parindent}{0.0cm}
     \bigskip\bigskip      {\Large {\bf Discussion}} \bigskip}
\def\speaker#1{{\bf #1:}\ }
\def\endDiscussion{}

\end{document}